\documentclass[12pt]{iopart}

\usepackage{graphicx}

\begin{document}

\title{Dimer XXZ Spin Ladders: Phase diagram and a Non-Trivial Antiferromagnetic Phase}

\author{Qi-Hui Chen, Long-Fei Guo and Peng Li\renewcommand{\thefootnote}{\arabic{footnote}} \footnotemark[1]}

\address{College of Physical
Science and Technology, Sichuan University, Chengdu 610064, China}
\ead{lipeng@scu.edu.cn}

\footnotetext[1]{Author to whom any correspondence should be addressed.}

\begin{abstract}
We study the dimer $XXZ$ spin model on two-leg ladders with isotropic
Heisenberg interactions on the rung and anisotropic $XXZ$ interactions along
the rail in an external field. Combining both analytical and numerical
methods, we set up the ground state phase diagram and investigate the
quantum phase transitions and the properties of rich phases, including the
full polarized, singlet dimer, Luttinger liquid, triplon solid, and a non-trivial
antiferromagnetic phases with gap. The analytical analyses based on solvable
effective Hamiltonians are presented for clear view of the phases and
transitions. Quantum Monte Carlo and exact diagonalization methods are
employed on finite system to verify the exact nature of the phases and
transitions. Of all the phases, we pay a special attention to the gapped
antiferromagnetic phase, which is disclosed to be a non-trivial one
that exhibits the time-reversal symmetry. We also discuss how our findings
could be detected in experiment in the light of ultracold atoms technology
advances.

\end{abstract}

\maketitle

\section{Introduction}
Ladders (or ladders-like) systems are of great interest in condensed matter
physics. On one hand, spin ladders materials with intriguing magnetic
properties have been found or synthesized \cite{Dagotto,Bouillot}. On the
other hand, due to the advances of cold atom technology, optical ladders
might be provided as one of the simplest accessible objects that exhibit
fascinating properties of quantum matters \cite{Jaksch,Greiner,Lmduan,Trotzky,folling1,Chen,sebby,crepin}. Apparently, real
compounds are not so flexible as the cold atoms in optical lattices in
exploring the underlying mechanism of theoretical models. Progresses in
these two fields could interplay and lead to a promising boom of new
findings.

In this work, we investigate a two-leg spin ladders in the strong anisotropic
limit. We construct the ground state phase diagram and investigate the
properties of rich phases, including the full polarized (FP), singlet dimer
(SD), Luttinger liquid (LL), triplon solid (TS), and non-trivial
antiferromagnet (non-trivial AF) ones. We set up the low-energy effective
models for describing different phases within the frame of bond operator
theory. Numerical methods including exact diagonalization (ED) and quantum
Monte Carlo (QMC) are employed to verify our new findings. We also discuss
how our interesting results could be related to experiment.

\section{The model}

The model we are concerned with is the dimer $XXZ$ spin-1/2 model on two-leg ladders
\begin{eqnarray}
H &=&\sum_{j}\sum_{l=1,2}\left[ J^{x}\left(S_{l,j}^{x}S_{l,j+1}^{x}+S_{l,j}^{y}S_{l,j+1}^{y}\right)
+J^{z}S_{l,j}^{z}S_{l,j+1}^{z}-hS_{l,j}^{z}\right]  \nonumber \\
&&+\sum_{j}K\mathbf{S}_{1,j}\cdot \mathbf{S}_{2,j}.  \label{H}
\end{eqnarray}
where $S_{l,j}^{\alpha }$ denotes the $S=1/2$ spin operator of $\alpha $ ($\alpha =x,y,z$) component at site $j$ ($j=1,2,...,L$) of $l$-th ($l=1,2$)
leg. $J^{x},J^{z}$ are the interactions strength along rail in transverse
and longitude directions respectively, $K$ is the Heisenberg interaction
strength along rung and will be set as the energy unit, $h$ denotes the
Zeeman field. For convenience, we consider the case of $J^{x}>0$ and $J^{z}>0 $, and a bipartite lattice without frustration, i.e. $L$ is even
when periodic boundary condition is imposed. We will focus on the strong
anisotropic cases, e.g. $J^{x}\rightarrow 0$ and $J^{z}\neq 0$, which are
not readily to be found in real compounds. Recent progress in the technology
of cold atoms are encouraging \cite{Chen}. The above model can be translated
to an identical hardcore bosonic $t$-$V$ model by a Matsubara-Matsuda mapping
\cite{Masubara}. Although some experiment on hardcore bosons have been
realized \cite{Paredes}, but the interaction terms in this model need more
sophisticated methods to realize \cite{Lmduan,Chen}. One may consider a
ladder-shaped optical lattice formed by standing wave lasers \cite{Pbhe}.
The superexchange mechanism could be mimicked by controlled tunneling of
cold bosonic or fermionic atoms with two internal states, such as $^{87}$Rb
in $|\uparrow \rangle =b_{\uparrow }^{\dag }|0\rangle =|F=1,m_{F}=1\rangle $
or $|\downarrow \rangle =b_{\downarrow }^{\dag }|0\rangle
=|F=1,m_{F}=-1\rangle $, where $|0\rangle $ is a vacuum. So that we
can define a spin on a site in a bilinear form $S^{\alpha }=\frac{1}{2}b^{\dag}\sigma^{\alpha }b$, where $\sigma^{\alpha },(\alpha=x,y,z)$, is a pauli matrix and $b^{\dag}=(b_{\uparrow}^{\dag},b_{\downarrow}^{\dag})$. In the Mott
insulating phase, the above model could be realized with almost arbitrary
model parameters \cite{Trotzky,Chen}.

Before solving the model, we introduce a bond operator representation that
will be very useful for elaborating the phenomena we found in this system.
There are four states on each rung, one singlet and three triplets,
\begin{eqnarray*}
|s\rangle _{j} &=&\frac{1}{\sqrt{2}}(|\uparrow \downarrow \rangle
_{j}-\left\vert \downarrow \uparrow \right\rangle _{j}), \\
|t_{+}\rangle _{j} &=&|\uparrow \uparrow \rangle _{j}, \\
|t_{0}\rangle _{j} &=&\frac{1}{\sqrt{2}}(|\uparrow \downarrow \rangle
_{j}+\left\vert \downarrow \uparrow \right\rangle _{j}), \\
|t_{-}\rangle _{j} &=&|\downarrow \downarrow \rangle _{j}.
\end{eqnarray*}
It is natural to introduce the bond operators $s_{j}^{\dag }$, $t_{+,j}^{\dag }$, $t_{0,j}^{\dag }$, $t_{-,j}^{\dag }$, which create the
singlet state and triplet states at $j$-th rung with the constraint, $s_{j}^{\dag }s_{j}+t_{+,j}^{\dag }t_{+,j}+t_{0,j}^{\dag
}t_{0,j}+t_{-,j}^{\dag }t_{-,j}=1$. Then the original Hamiltonian Eq.(\ref{H}) can be rewritten in terms of these bond operators as
\begin{eqnarray}
H &=&\frac{K}{4}\sum_{j}(-3s_{j}^{\dag }s_{j}+t_{+,j}^{\dag
}t_{+,j}+t_{0,j}^{\dag }t_{0,j}+t_{-,j}^{\dag }t_{-,j})+  \nonumber \\
&&\frac{J^{x}}{2}\sum_{j}[-s_{j}^{\dag }s_{j+1}^{\dag
}(t_{+,j}t_{-,j+1}+t_{-,j}t_{+,j+1})+s_{j}^{\dag
}s_{j+1}(t_{+,j}t_{+,j+1}^{\dag }+t_{-,j}t_{-,j+1}^{\dag })+  \nonumber \\
&&t_{0,j}t_{0,j+1}(t_{+,j}^{\dag }t_{-,j+1}^{\dag }+t_{-,j}^{\dag
}t_{+,j}^{\dag })+t_{0,j}t_{0,j+1}^{\dag }(t_{+,j}^{\dag
}t_{+,j+1}+t_{-,j}^{\dag }t_{-,j+1})+h.c]+  \nonumber \\
&&\frac{J^{z}}{2}\sum_{j}[(s_{j}^{\dag }s_{j+1}^{\dag
}t_{0,j}t_{0,j+1}+s_{j}^{\dag }s_{j+1}t_{0,j}t_{0,j+1}^{\dag }+h.c)+  \nonumber
\\
&&(n_{+,j}-n_{-,j})(n_{+,j+1}-n_{-,j+1})]-h\sum_{j}(n_{+,j}-n_{-,j}),
\label{Hb}
\end{eqnarray}
where, $n_{\pm ,j}=t_{\pm ,j}^{\dag }t_{\pm ,j}$ are the particle number
operators of $t_{\pm ,j}$.

\section{Phase diagram and properties of the phases}

We have employed both stochastic series expansion (SSE) quantum Monte Carlo
(QMC) \cite{sandvik1,sandvik2} and exact diagonalization methods to
investigate numerically the system's low energy properties. Analytical
methods have also been applied, including the bond-operator mean-field
(BOMF) theory and bosonization method for a comparison and comprehension of
the underline physics. In this work, the SSE QMC simulation is performed on
a ladder of length up to $L=160$. The temperature is taken as $T=3/L$ and
periodic boundary condition is imposed. Thus the lowest temperature $T=0.01875$ is reached, which is sufficient to obtain the ground state
observables. We present two slices of phase diagrams to exemplify our
results: (i) $J^{x}=0$ (Figure 1); (ii) $J^{x}=0.35$ (Figure 2). The
properties of the ground states and the low energy excitations are
elaborated as follows.

\begin{figure}[tbp]
\includegraphics[width=0.9\textwidth,trim=-200 80 0 0]{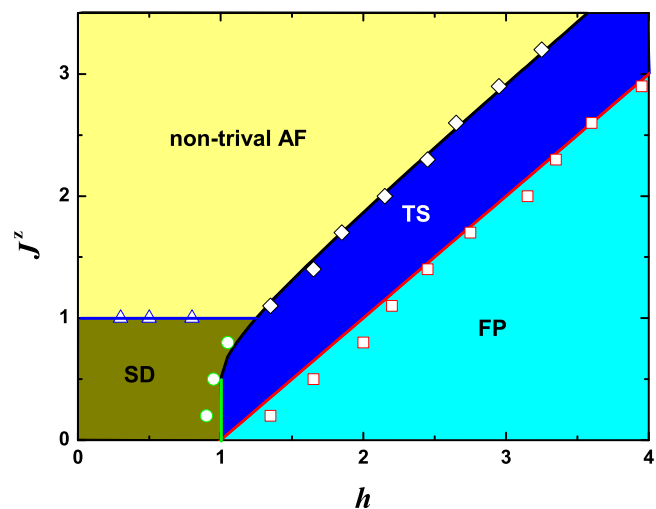} \center
\caption{(Color online) The phase diagram with $J^{x}=0$. The
solid lines and scattering lines come from bond operator theory and quantum
Monte Carlo, respectively. The phase diagram contains an antiferromagnet
(AF), a singlet dimer (SD), a triplon solid (TS) and fully polarized state
(FP).}
\end{figure}

\begin{figure}[tbp]
\includegraphics[width=1.0\textwidth,trim=-300 80 0 0]{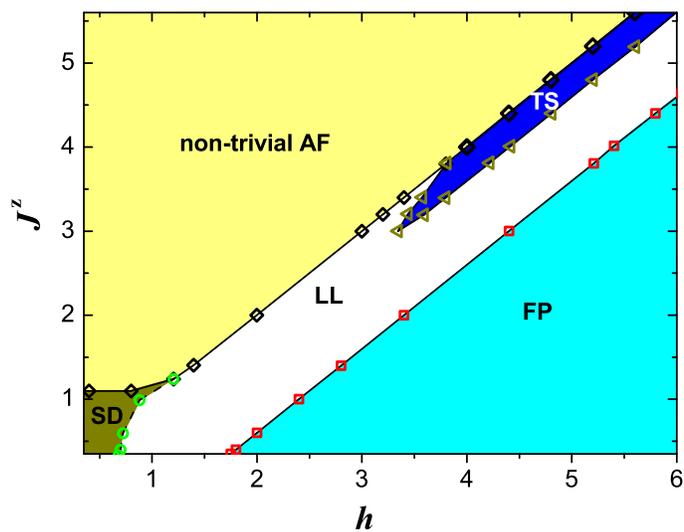}\newline
\caption{(Color online) Ground state phase diagram with $J^{x}=0.35$. The scatter lines (Black) are QMC results. There are four
orders which are antiferromagnetic (AF), full polarized (FP), and Luttinger
liquid (LL).}
\end{figure}
\subsection{Full polarized and singlet dimer phases}

First, let us see two simplest limits. The first is the full polarized (FP)
state in large enough field $h$. The second occurs at $J^{x}=J^{z}=0$, where
the Hamiltonian (\ref{H}) reduces to an array of disjoint dimers. In the
second limit, the ground state and excitations are determined by individual
dimers. The energies of singlet and triplets are $E(s_{j})=-3K/4$, $E(t_{+,j})=K/4-h$, $E(t_{0,j})=K/4$, $E(t_{-,j})=K/4+h$. When $h=0$, the
ground state is an array of singlets (we call it singlet dimer (SD) phase)
with an excitation gap $\Delta _{1}=K$. With the applied field $h$
increasing, the system's ground state experiences a transition from SD to FP
state at $h=\Delta _{1}$.
\subsection{Triplon solid}

If we fix $J^{x}=0$ and raise $J^{z}>0$ ($<1$), a triplon solid (TS) phase
intervenes between SD and FP phases (Figure 1). The TS state is a paving of
singlets and triplets on the ladders alternatively. To detect the TS order
in a general case, one can define a triplon creation operator $d_{i}^{\dag}$
acting in spin's Hilbert space
\begin{equation}
d_{j}^{\dag }=\frac{1}{\sqrt{2}}(S_{1,j}^{+}-S_{2,j}^{+}).  \label{b}
\end{equation}
We have $d_{j}^{\dag }|t_{-}\rangle _{j}=|s\rangle _{j}$, $d_{j}^{\dag
}|s\rangle _{j}=|t_{+}\rangle _{j}$, and $d_{j}^{\dag }|t_{+}\rangle _{j}=0$
. It is worth noting that $d_{j}^{\dag }$ is not a perfect hardcore boson \cite{NgLee}. But with appropriate model parameters, the hardcore condition
could be well fulfilled. The diagonal real space correlation function of
triplon is defined as
\begin{equation}
C^{zz}(r)=\frac{1}{L}\sum_{j}\langle t_{+,j}^{\dag }t_{+,j}t_{+,j+r}^{\dag
}t_{+,j+r}\rangle ,
\end{equation}
where the thermodynamic limit $L\rightarrow \infty $ is taken. The static
structure factor for detecting TS order is given by
\begin{equation}
S(Q)=\frac{1}{L}\sum_{r}C^{zz}(r)e^{\mathbf{i}Qr}.  \label{spi}
\end{equation}

From the point of view of renormalization, the $|t_{0}\rangle _{j}$ and $
|t_{-}\rangle _{j}$ states could be projected out from Hilbert space in a
moderate strong external field, which breaks the time-reversal symmetry (TRS).
Thus we get a reduced constraint $s_{j}^{\dag }s_{j}+t_{+,j}^{\dag
}t_{+,j}=1 $, so that the Hamiltonian (\ref{Hb}) at low energy sector can be
replaced by
\begin{equation}
H_{1}=\frac{J^{z}}{2}\sum_{j}\tau _{j}^{z}\tau _{j+1}^{z}-\left( h-K-\frac{J^{z}}{2}\right) \sum_{j}\tau _{j}^{z}+L\left( \frac{J^{z}}{8}-\frac{K}{4}-\frac{h}{2}\right) ,  \label{TS}
\end{equation}
where we have defined the pseudo-spin operators in a Schwinger boson
representation, $\tau _{j}^{\dag }=t_{+,j}^{\dag }s_{j},\tau
_{j}^{-}=t_{+,j}s_{j}^{\dag },\tau _{j}^{z}=(t_{+,j}^{\dag
}t_{+,j}-s_{j}^{\dag }s_{j})/2$. The effective Hamiltonian (\ref{TS}) is
nothing but a "classical Ising model in a field". The TS
phase is a consequence of the first antiferromagnetic interaction term. If
the second term prevails the first one, SD and FP phases could be reached if
$\left( h-K-\frac{J^{z}}{2}\right) <0$ and $\left( h-K-\frac{J^{z}}{2}
\right) >0$ respectively. We also get the phase boundaries by comparing the
ground state energies of the three states: (i) $h=K$ for the boundary
between SD and TS and (ii) $J^{z}=h-K$ between TS and FP. It is worth to
point out that the effective Hamiltonian (\ref{TS}) also gives a good
description for TS and FP phase in the phase diagram where $J^{z}>1$. The
consistency between the analytical and numerical results are quite well, as
shown in Figure 3.
\begin{figure}[tbp]
\includegraphics[width=0.9\textwidth,trim=-300 80 0 0]{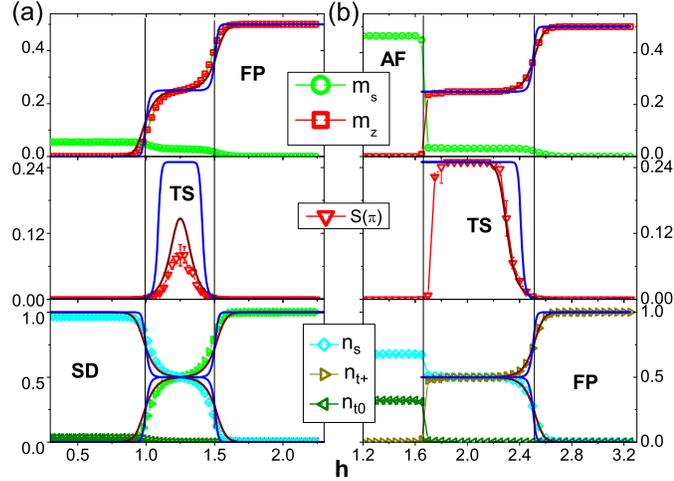}\newline
\center
\caption{(Color online) The order parameters defined by Eq.(\protect\ref{spi}) and
(\protect\ref{msz}) and the densities of state $|s\rangle ,|t_{+}\rangle
,|t_{0}\rangle $ per rung. The selected parameters are $J^{x}=0,J^{z}=0.5$
in (a) and $J^{x}=0,J^{z}=1.5$ in (b). The scattered lines represent QMC
results with $L=100$. The wine-colored solid lines are results obtained
from bond operator theory with $L=100$ (blue lines are for $L=300$).}
\end{figure}
\subsection{Luttinger liquid}

If $J^{x}\neq 0$, the area of TS phase shrinks to the top right corner of
the phase diagram (Figure 2) and there emerges a vast gapless phase called
Luttinger liquid (LL). In isotropic case, i.e., $J^{x}=J^{z}$, there have
been a large amount of literatures devoted to the study of the ground state
and low excitations \cite{Sczhang,Laflorencie,Giamarchi1999}. Here we focus
on the anisotropic case, $J^{z}\neq J^{x}$, and answer the question whether
there is still a region described by LL, which, to our knowledge, is not
been discussed before. We give a positive answer here.

Interestingly, the low energy properties in the LL phase are also governed
by the states $|s\rangle _{j}$ and $|t_{+}\rangle _{j}$. But the effective
Hamiltonian includes a part of "quantum fluctuations" besides the
"classical" part, $H_{1}$,
\begin{equation}
H_{2}=H_{1}+\sum_{j}[\frac{J^{x}}{2}(\tau _{j}^{+}\tau _{j+1}^{-}+h.c)].
\label{LL}
\end{equation}
This model is exactly the form of the $XXZ$ chain whose properties are well
known owing to many reliable methods, such as Bethe Ansatz and bosonization.
We know that its ground state is antiferromagnet or LL when $J^{z}/J^{x}>2$
or $J^{z}/J^{x}<2$ respectively \cite{Yang}. In LL, both diagonal and
off-diagonal triplon correlations are important. The former is defined in (\ref{spi}). The latter can be defined as
\begin{equation}
C^{+-}(r)=\frac{1}{L}\sum_{j}\langle d_{j}^{\dag }d_{j+r}\rangle .
\end{equation}
From the effective model (\ref{LL}), we can get the equal-time spin
correlation by the standard Abelian bosonization techniques \cite{Hikihara},
\begin{equation}
C^{zz}(r)=\langle \tau _{j+r}^{z}\tau _{j}^{z}\rangle =m_{z}^{2}-\frac{c_{1}
}{r^{2}}+\frac{c_{2}}{r^{2K_{L}}}\cos (2\pi m_{z}r),
\end{equation}
\begin{equation}
C^{+-}(r)=\langle \tau _{j+r}^{+}\tau _{j}^{-}\rangle =c_{3}\frac{(-1)^{r}}{r^{1/2K_{L}}}-c_{4}\frac{(-1)^{r}}{r^{2K_{L}+1/2K_{L}}}\cos (2\pi m_{z}r),
\label{cpm}
\end{equation}
where $r$ denotes the lattice distance, $c_{1},c_{2},c_{3},c_{4}$ are
constants, $m_{z}=(1/2N)\sum_{j}\langle (S_{1,j}^{z}+S_{2,j}^{z})\rangle $
is the uniform magnetization, $K_{L}$ is the so-called LL parameter with
typical value less than $0.5$ that governs the spin correlations at long
wavelength limit. For a comparison, we perform the QMC simulation on the
original Hamiltonian (\ref{H}). We found the spin ladder is sensitive to the
value of $J^{x}$ in TS region. However, comparing with spin $XXZ$ chain, the
TS and LL still exist in the phase diagram of the spin ladder only with
quantitative changes of phase boundaries. The densities of four rung states $|s\rangle ,|n_{t+}\rangle ,|n_{t0}\rangle ,|n_{t-}\rangle $ shown in Figure
4(b) confirm the goodness of the effective model (\ref{LL}). From the nodes
of structure factor $S(\pi )$ in Figure 4(a), we can see a clear phase
transition from TS to LL. The long range diagonal order in TS also could be
seen in the correlation function $C^{zz}(r)$ of Figure 4(d). In Figure 4(c)
we show the power law decay behavior of correlation $C^{+-}(r)$ in LL
region, which coincides with the relation (\ref{cpm}) perfectly.
\begin{figure}[tbp]
\includegraphics[width=0.8\textwidth,trim=-300 80 0 0]{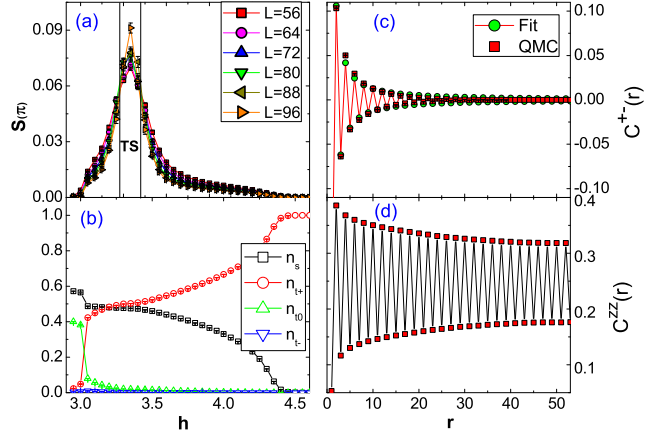}\newline
\caption{(Color online) (a) The structure factor $S(\protect\pi )$ in $h=3.8$
obtained by QMC in different lattice sizes. (b) The densities of four states
in each rung. (c) The correlation function $C^{+-}(r)$ where the square
scattered line is obtained from QMC and the circular scattered line comes
from fitting Eq.(\protect\ref{cpm}). (d) The correlation function $C^{zz}(r)$
with long range order in $h=3.3$. Here, the selected parameters are $J^{x}=0.35,J^{z}=3.0$.}
\end{figure}

\subsection{Non-trivial antiferromagnetic phase}

Now we turn to the last phase. First, we focus on the thermodynamic limit,
and the case of $J^{x}=0$ (see Figure 1), since the qualitative properties do
not change if $J^{x}$ is not large enough. We show the ground state phase is
a gapped Mott insulator of triplon that can be understood quite good by an
effective Ising model in a transverse field (T-Ising model). Then we will
work on finite system to figure out the low excitations of this phase. We will show the lowest two states for finite systems are nearly degenerate and tend to be maximally entangled with $J^z$ increasing.
\subsubsection{Two channels}

When $h$ is small and $J^{z}$ is large enough, the low energy sector of the
system splits into two channels. The first one (channel A) is a classical
Ising Hamiltonian
\begin{equation}
H_{A}=2J^{z}\sum_{j}\eta _{j}^{z}\eta _{j+1}^{z}-2h\sum_{j}\eta _{j}^{z}+
\frac{LK}{4},  \label{nt+-}
\end{equation}
with $\eta _{j}^{z}=\frac{1}{2}(t_{+,j}^{\dag }t_{+,j}-t_{-,j}^{\dag
}t_{-,j})$ and constraint $t_{+,j}^{\dag }t_{+,j}+t_{-,j}^{\dag }t_{-,j}=1$.
When $h=0$, the gound state is an alternative paving of $|t_{+}\rangle _{j}$
and $|t_{-}\rangle _{j}$ states along the ladders. Its lowest energy per
rung reads
\begin{equation}
E_{0,A}/L=\frac{K}{4}-\frac{J^{z}}{2}.  \label{E0nt+-}
\end{equation}

The second one (channel B) is obtained by introducing a set of pseudo-spin
operators in another Schwinger boson representation, $\lambda
_{j}^{+}=s_{j}^{\dag }t_{0,j}$, $\lambda _{j}^{-}=s_{j}t_{0,j}^{\dag }$, $\lambda _{j}^{z}=(s_{j}^{\dag }s_{j}-t_{0,j}^{\dag }t_{0,j})/2$, with
constraint $s_{j}^{\dag }s_{j}+t_{0,j}^{\dag }t_{0,j}=1$, which reads
\begin{equation}
H_{B}=\sum_{j}[2J^{z}\lambda _{j}^{x}\lambda _{j+1}^{x}-K\lambda
_{j}^{z}-K/4].  \label{TIsing}
\end{equation}
This is exactly the T-Ising model, whose properties are also well-known. One
can solve it exactly by applying the Jordan-Wigner transformation \cite{Kitaev}, and get the lowest energy per
rung,
\begin{equation}
E_{0,B}/L=-\frac{K}{4}-\frac{J^{z}(J^{z}+K)}{\pi }E\left( \frac{4J^{z}K}{(J^{z}+K)^{2}}\right) ,
\end{equation}
where $E(x)$ is the complete elliptic integral of the second kind. Because
the field $h$ is coupled with $t_{i,+}$ and $t_{i,-}$, it does not
appear in channel B. In fact, this is the reason why the phase boundary
between SD and AF phases is a straight line (see Figure 1 and 2) that almost
does not change with the field increasing for a given $J^{x}$ until runs
into another phase. In the following, combining numerical approaches, we
emphasize several important consequences of the above effective
Hamiltonians. It is easy to see that the difference of the lowest energies
between channels A and B is
\begin{equation}
\Delta =E_{0,A}/L-E_{0,B}/L>K/2.  \label{Delta}
\end{equation}

So we see that channel B truly reflects the ground state of the original
ladders system. In the whole parameter region, channel B gives quite good
description of the ground state energy of original system. The goodness of
effective $H_{B}$ for the ground state has been confirmed by QMC simulations
on the original ladders system shown in Figure 3, 4, and 5.
\subsubsection{Quantum phase transition in thermodynamic limit}

In thermodynamic limit and when $J^{z}>K$, the effective Hamiltonian $H_{B}$
exhibits an AF phase with double degeneracy,
\begin{eqnarray}
|\psi _{1}^{AF}\rangle  &=&\prod_{j}\otimes \left( |s\rangle
_{j}+(-1)^{j}|t_{0}\rangle _{j}\right) ,  \label{AF1} \\
|\psi _{2}^{AF}\rangle  &=&\prod_{j}\otimes \left( |s\rangle
_{j}-(-1)^{j}|t_{0}\rangle _{j}\right) .  \label{AF2}
\end{eqnarray}
This result needs an assumption of spontaneous symmetry breaking
nevertheless. We will address this issue later. The staggered magnetization
of this phase can be detected by the order parameter
\begin{equation}
m_{s}=\frac{1}{2L}\sum_{i}(-1)^{i}\langle (S_{1,j}^{z}-S_{2,j}^{z})\rangle .
\label{msz}
\end{equation}
There is a quantum phase transition between the SD and AF phases at the
gapless point $J^{z}=K$, which can be worked out from the divergence point
of the second-order derivative of $E_{0,B}$ on $J^{z}$, as well as the
first-order derivative of the weight of singlets
\begin{equation}
n_{s}\equiv \frac{1}{L}\sum_{j}\langle s_{j}^{\dag }s_{j}\rangle =\frac{1}{2\pi }\int_{-\pi }^{\pi }dk\frac{(J^{z}\sin k)^{2}}{2\omega (k)\left( \omega
(k)+J^{z}\cos k-K\right) },
\end{equation}
where
\begin{equation}
\omega (k)=\sqrt{(J^{z}\cos k-K)^{2}+(J^{z}\sin k)^{2}}  \label{WB}
\end{equation}
is the lowest excitation spectrum of $H_{B}$. The staggered magnetization $m_{s}$ is deduced from the correlation function $\langle \lambda
_{i}^{x}\lambda _{j}^{x}\rangle $ formally, which is too tedious to be
presented here. The smooth curve in Figure 5(a) shows $m_{s}$ that is
obtained by the correlation function with $\max |i-j|=160$. Comparing it
with QMC results, we confirm the validity of the effective Hamiltonian $H_{B}
$. The phase transition between those two states is of second order, which
can be seen from the staggered magnetic susceptibility $\chi _{s}$ shown in
Figure 5(b).
\begin{figure}[tbp]
\includegraphics[width=0.8\textwidth,trim=-300 80 0 0]{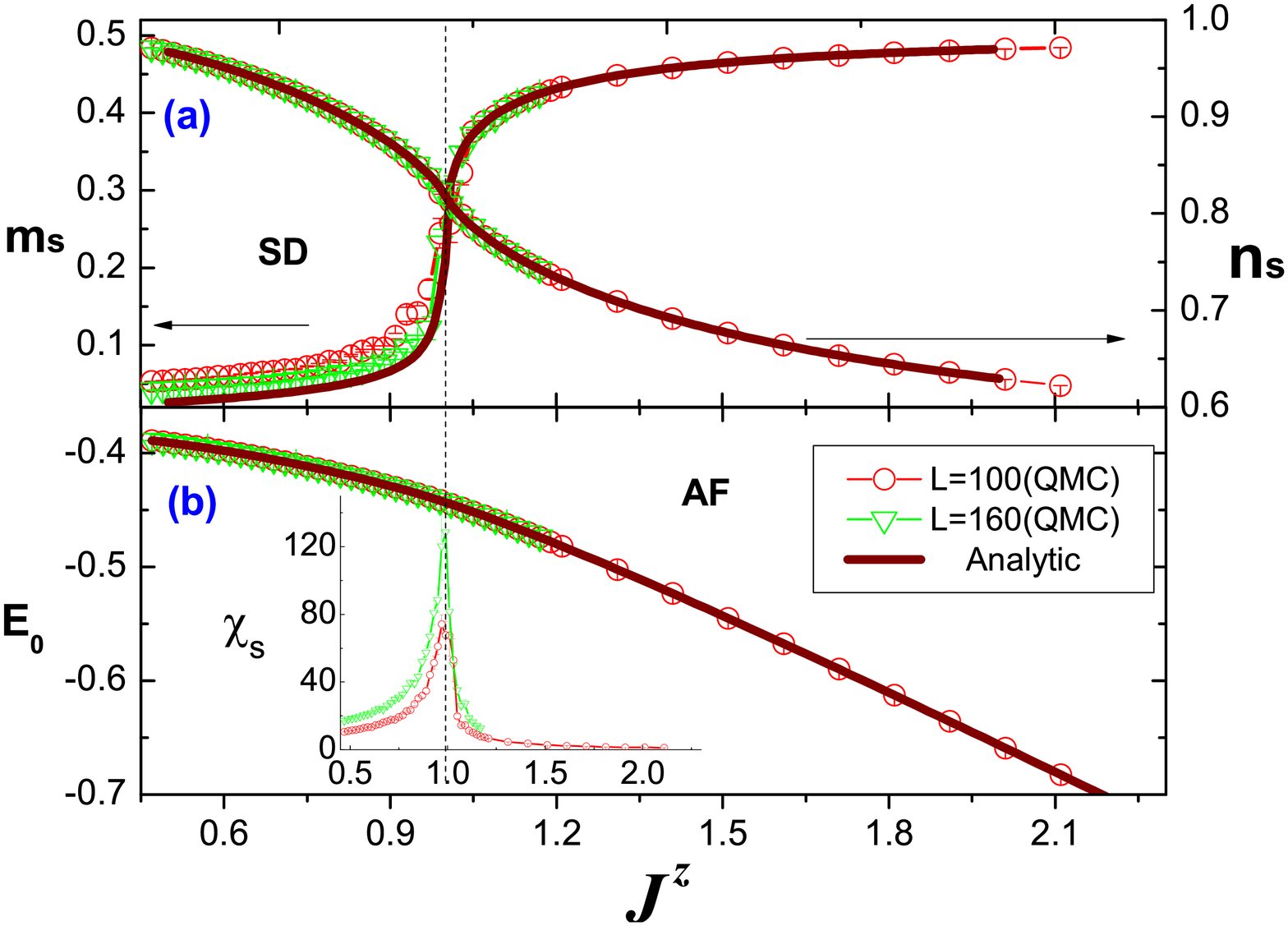}\newline
\caption{(Color online) (a) The staggered magnetization and the density of
singlet state per rung. (b) The energy per site and the inserted one is
staggered magnetic susceptibility. Here the scattered lines are obtained
from QMC, the solid line are analytic results of effective Hamiltonian (\protect\ref{TIsing}). The QMC is simulated in a ladder with $J^{x}=0$, $h=0.5$.}
\end{figure}
\subsubsection{Exact diagonalization of finite systems}

Is the effective Hamiltonian $H_{B}$ good enough for low excitations? How do
the translational and TRS breaking happen in the AF phase
for the original ladders system? To answer these questions, we performed an
exact diagonalization on the original system up to $L=10$ (i.e. $20$ sites)
to reveal the aspect of the original system that may be ignored. We found,
although the effective $H_{B}$ reflects the ground state of the original
system perfectly, the low excitations above the ground state is largely
restricted by channel A. The excitation spectrum (\ref{WB}) of channel B
gives an energy gap above its ground state,
\begin{equation}
\Delta _{gap}^{B}=\left\vert J^{z}-K\right\vert .
\end{equation}
But the true gap of the original system should not exceed $\Delta $ (see (\ref{Delta})), so in large $J_{z}$ we have
\begin{equation}
\Delta _{gap}=\min (\Delta ,\Delta _{gap}^{B}).
\label{gap}
\end{equation}
\begin{figure}[tbp]
\begin{center}
\includegraphics[width=4.3915in]{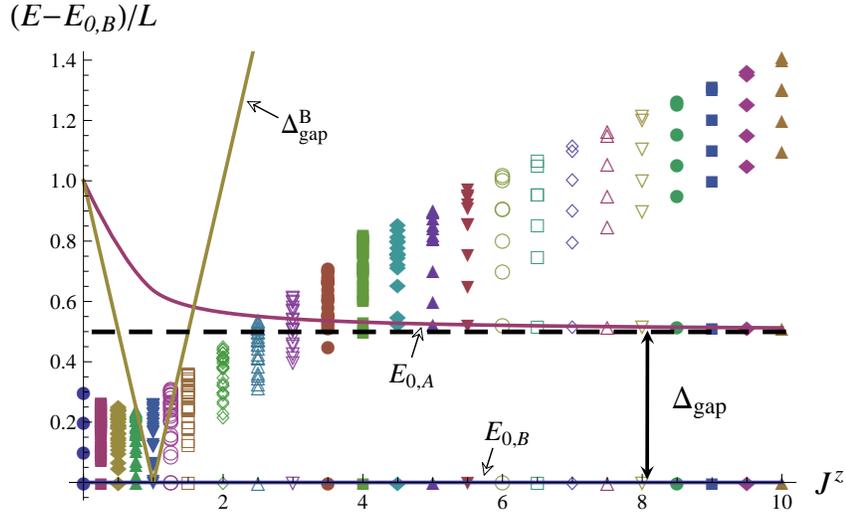}
\end{center}
\caption{(Color online) Discrete energy levels for a ladder with length $L=10$ (labelled by colored markers) obtained by exact diagonalization and
analytical results for a ladder with $L\rightarrow \infty $ (shown by lines)
obtained by effective theories. The dashed line represents the true gap at large $J_{z}$ limit. Here $J^x=0$ and $K=1$ are taken. Please see more details in
the text.}
\end{figure}
\begin{figure}[tbp]
\begin{center}
\includegraphics[width=4.3915in]{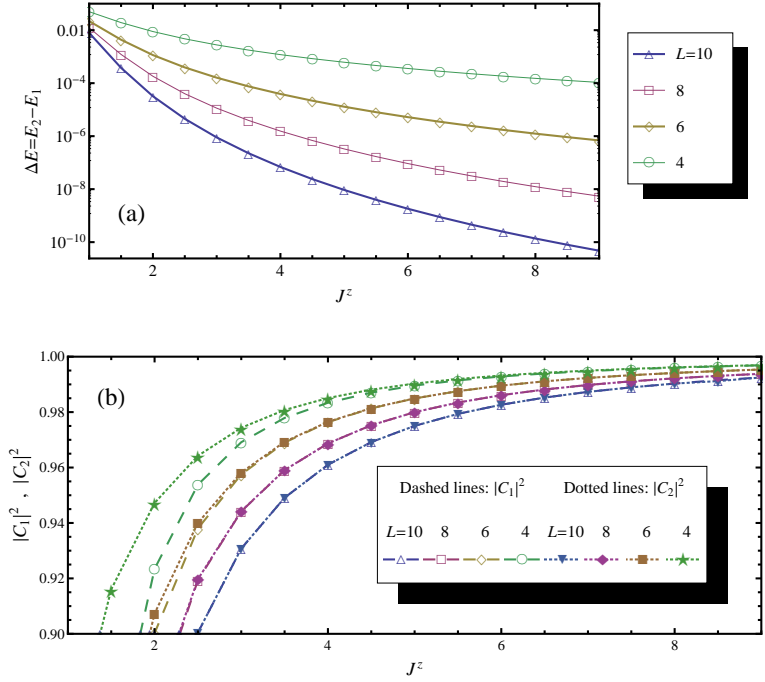}
\end{center}
\caption{(Color online) (a) The energy difference, $\Delta E=E_{2}-E_{1}$, of the lowest two states, $\left\vert \phi _{1}\right\rangle$ and $\left\vert \phi _{2}\right\rangle$. (b) The percentages, $|C_{1}|^2$ and $|C_{2}|^2$, of dominant parts in the lowest two states. Please see more in the text.}
\end{figure}
The numerical result for $J^x=0$ is exemplified in Figure 6. We see the excitations of the
ladder system is much different from the effective $H_{B}$. In large $J^{z}$, we observe no extra energy levels lying between $E_{0,A}$ and $E_{0,B}$,
which ensures a true gap greater than $K/2$. Another observation is the
lowest two levels of finite systems are not exactly degenerate and not even
the ones with AF order, $|\psi _{1}^{AF}\rangle $ and $|\psi_{2}^{AF}\rangle $. Instead, they are
\begin{eqnarray}
\left\vert \phi _{1}\right\rangle  &=&\frac{C_{1}}{\sqrt{2}}\left( |\psi _{1}^{AF
}\rangle +|\psi _{2}^{AF}\rangle \right) +res., \\
\left\vert \phi _{2}\right\rangle  &=&\frac{C_{2}}{\sqrt{2}}\left( |\psi _{1}^{AF
}\rangle -|\psi _{2}^{AF}\rangle \right) +res.,
\end{eqnarray}
with dominant parts prevail the residuals. The numerical results are shown in Figure 7. We observed that $\left\vert \phi _{1}\right\rangle $ has slightly lower
energy than $\left\vert \phi _{2}\right\rangle $, but their energy
difference $\Delta E=E_{2}-E_{1}$ rapidly reduces to zero with $J_{z}$ increasing.
And, at the same time, the two states approach the so-called GHZ states \cite{GHZ} asymptotically,
\begin{eqnarray}
\left\vert \phi _{1}\right\rangle  &\approx &\frac{1}{\sqrt{2}}\left(
|\psi _{1}^{AF}\rangle +|\psi _{2}^{AF}\rangle \right) , \\
\left\vert \phi _{2}\right\rangle  &\approx &\frac{1}{\sqrt{2}}\left(
|\psi _{1}^{AF}\rangle -|\psi _{2}^{AF}\rangle \right) .
\end{eqnarray}

The result implies the entangled GHZ states could be purified if we increase $J^z$ adiabatically for finite system. For long enough system, spontaneous symmetry breaking may occur and the entanglement will vanish. Nonetheless, the two GHZ states do not exhibit a non-zero order parameter $m_{s}$, while they do have
the same antiferromagnetic spin correlation functions as $|\psi _{1}^{AF}\rangle $
and $|\psi _{2}^{AF}\rangle $. That is the reason why we name it a non-trivial AF phase. One can check that $\left\vert \phi _{1}\right\rangle $ and $\left\vert
\phi _{2}\right\rangle $ are eigenstates of the time reversal operator
\begin{equation}
\Theta =\prod\limits_{j,l}\otimes \left( -\mathbf{i}\sigma _{l,j}^{y}\right)
\kappa ,
\end{equation}
where $\kappa $\ is the complex conjugation, if noticing that
\begin{eqnarray}
\Theta s_{j}^{\dag }\Theta ^{\dag } &=&s_{j}^{\dag }, \\
\Theta t_{0,j}^{\dag }\Theta ^{\dag } &=&-t_{0,j}^{\dag }.
\end{eqnarray}
Thus $\left\vert \phi _{1}\right\rangle $ and $\left\vert \phi
_{2}\right\rangle $ preserve the exact TRS as the same as
the original Hamiltonian (\ref{H}). From the point of view of the effective $H_{B}$, the assumption of spontaneous symmetry breaking in thermodynamic limit is based an unpredictable disturbance from environment \cite{Kitaev}. But here we have one favorable factor to avoid this - the destructive field should couple to local variables, $\lambda_{j}^{x}=(s_{j}^{\dag }t_{0,j}+s_{j}t_{0,j}^{\dag })/2$, delicately since the system with true gap is antiferromangetic. Thus we think there is the chance to realize such entangled states that do not break the TRS.
\section{Discussion}

The dimer $XXZ$ spin system discussed in this paper might be realized in a Mott
insulating phase of cold bosonic atoms in near future \cite{Trotzky,Chen}.
We can estimate the typical energy scales. For $^{87}$Rb atoms with a
lattice constant $\pi /k_{x}\simeq 2\pi /k_{y}\sim 426$ nm and about $10^{5}$
atoms in a Bose-Einstein condensate, we can chose $t^{2}/(\hbar U)\sim 0.1$
kHz (corresponding to a time scale of 10 ms) with a conservative choice of $U\sim $ $2$ kHz and $(t/U)^{2}\sim 1/20$, where $t$ is the hopping strength of cold atoms between two nearest neighbor minimums of laser potential and $U$ is on site interaction originating from the $s$ wave scattering. These energy scales are clearly
compatible with current experiments \cite{Greiner} and make the system in a
Mott insulating area. In experiment, the density of condensates in momentum
space $\langle \tilde{n}_{\pi }\rangle $ can be measured by noise
correlations which can be linked to spin-spin correlations \cite{altman,folling,Scarola}. We can use Bragg scattering of light, which gives
rise to the spin structure factor, to detect $S(\pi )$ \cite{cor}. An
alternative technique for imaging spin states in optical lattices has been
put forward \cite{dou}. Thus, all the phases discussed in this paper are
detectable in experiment.

\section{Summary}

In summary, combining analytical and numerical methods, we have investigated
the ground state phase diagrams and low excitations of the dimer $XXZ$ spin
ladder system. We show that most features of the phases could be understood
within the frame of bond operator theory and have proven it by using quantum
Monte Carlo method. We present the rich ground state phase diagram which can
be detected in optical lattice by future experiment.
\section{Acknowledgments}

This work was supported by the NSFC under grants No. 11074177, SRF for ROCS
SEM (20111139-10-2)

\end{document}